\documentclass[]{rnaastex}

\usepackage{graphicx}
\usepackage[suffix=]{epstopdf}
\usepackage{natbib}
\usepackage{amsmath}
\usepackage{xspace}
\usepackage[overload]{textcase}

\usepackage{hyperref}

\begin{document}

\newcommand{\kepler}{\emph{Kepler}}
\newcommand{\spitzer}{{\it Spitzer}}
\newcommand{\ktwo}{{\it K2}}

\title{PREDICTING THE ORBIT OF TRAPPIST-1\MakeLowercase{i}}

\correspondingauthor{David Kipping}
\email{dkipping@astro.columbia.edu}

\author{David Kipping}
\affiliation{Department of Astronomy, Columbia University, 550 W 120th Street, New York NY}


\section{} 

The TRAPPIST-1 system provides an exquisite laboratory for advancing our
understanding exoplanetary atmospheres, compositions, dynamics and
architectures. A remarkable aspect of the TRAPPIST-1 is that it represents the
longest known resonance chain \citep{luger:2017}, where all seven planets share
near mean motion resonances (MMR) with their neighbors such that

\begin{align}
(j+k) P_x &= j P_{x+1},
\end{align}

where $P_x$ is the period of the $x^{\mathrm{th}}$ planet such that
$P_x < P_{x+1}$, $k$ and $j$ are integers describing the resonance order and
spacing respectively. For example, planets b and c are close to the 8:5
resonance, which is a $3^{\mathrm{rd}}$ order ($k=3$) resonance closely packed
with $j=5$. Because the MMRs are not exact, the location of conjunction
circulates with a period commonly called the ``super-period''
\citep{cochran:2011,lithwick:2012} with

\begin{align}
P_{\mathrm{super}} &= \frac{1}{|\frac{j+k}{P_{x+1}} - \frac{j}{P_x}|}.
\end{align}

More than this, neighboring triples reside in Laplace-like resonances
\citep{luger:2017}, characterized by

\begin{align}
\frac{p}{P_x} - \frac{p+q}{P_{x+1}} + \frac{q}{P_{x+2}} &\simeq 0,
\end{align}

where $p$ and $q$ are integers describing the resonances. As a result of the
Laplace-like resonances, the outer four pairs share the same super-period, with
the inner two having a integer ratio of this global super-period.

Although the period of 1h is now precisely known, the original discovery
of this outer planet was based upon just a single transit observation
\citep{gillon:2017}. Without repeated events, the period was only very
crudely constrained at the time, using the transit duration,
to $P_h = 20_{-6}^{+15}$\,days. After the discovery of new transits of
1h using \ktwo, the period was locked down to $18.767$\,days by
\citet{luger:2017}. However, the authors noted that this period could have
been predicted to several decimal places even before the \ktwo\ observations
using a resonance chain argument.

Assuming that planet 1h is also in a Laplacian resonance with planets
1f and 1g, and that $p$ and $q$ are in the range 1 to 3 (as the others
are), then six possible periods result:
i)   $(p,q)=(1,3) \to P_h = 13.941$\,d;
ii)  $(p,q)=(1,2) \to P_h = 14.899$\,d;
iii) $(p,q)=(2,3) \to P_h = 15.998$\,d;
iv)  $(p,q)=(1,1) \to P_h = 18.766$\,d;
v)   $(p,q)=(3,2) \to P_h = 25.345$\,d; and
vi)  $(p,q)=(2,1) \to P_h = 39.026$\,d. Using the transit duration, one
could potentially narrow this choice of six down, although in practise
almost all of these are compatible with the observed duration.

\clearpage

In this work we highlight that an additional, but unproven, way of choosing
between these periods is the generalized Titius-Bode (TB) law. The periods
of the inner six TRAPPIST-1 planets fall closely onto a power-law relation
given by 

\begin{align}
P_x &= P_1 \alpha^{x-1}.
\end{align}

Although originally posed for the Solar System planets, \citet{bovaird:2013}
showed that for \kepler\ 4+ planet systems, the TB law also holds. This led
the authors to predict the presence of 141 new exoplanets using the TB
relation. Taking a subset of 97 of these predictions, \citet{huang:2014}
found five new planets, but generally concluded the predictive power of
the law appeared questionable.

\begin{figure*}
\begin{center}
\includegraphics[width=15.5cm,angle=0,clip=true]{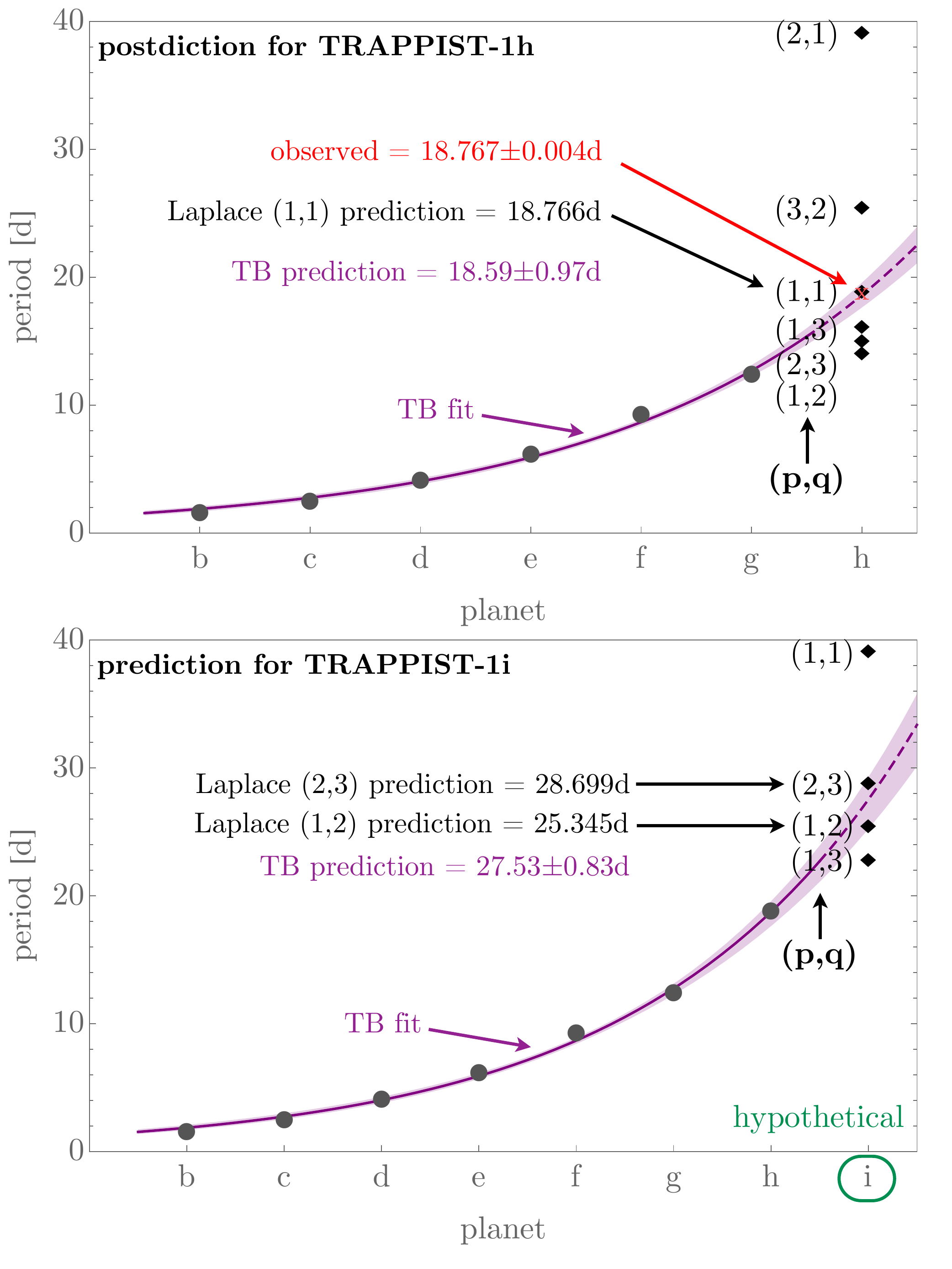}
\caption{\emph{
Upper: Postdiction for the orbital period of TRAPPIST-1h using the generalized
TB law argument of \citet{bovaird:2013} combined with the resonant chain
argument of \citet{luger:2017}. A unique and highly precise period is
forecasted, which is ultimately almost exactly the observed value. Lower:
Applying the same logic to predict the period of a hypothetical TRAPPIST-1i,
where we find two candidate periods satisfying both the TB and Laplacian
arguments.
}}
\label{fig:1}
\end{center}
\end{figure*}

\citet{hayes:1998} showed that the TB law is a natural outcome for multiplanet
systems satisfying the condition that they are Hill stable. TRAPPIST-1 is an
example of a System with Tightly-spaced Inner Planets (STIP) such that the
planets are dynamically packed within a compact region, and thus the argument
of \citet{hayes:1998} would indiciate that the TB law may be a natural
by-product of such an arrangement. Applying the TB law to the inner six
planets indeed reveals an excellent fit (as shown in Figure~\ref{fig:1}), with
an extrapolated seventh planet appearing at $18.59 \pm 0.97$\,d. Comparing
to the six possible resonant periods for 1h discussed earlier, the TB +
Laplacian arguments together would predict a single unique period of
$18.766$\,d, which is indeed bang on the observed period of $18.767 \pm
0.004$\,d \citep{luger:2017}.

A postdiction is never as impactful as a prediction. And so, we take the next
step and use this logic to predict the orbital period of TRAPPIST-1i. To be
clear, there is presently no literature suggesting an eighth planet but if it
should exist, we can predict its period. We first note that the TB law
predicts a period for 1i of $27.53 \pm 0.83$\,d. Exploring the possible
combinations of $p$ and $q$ below 3, we find five possible periods of
i)   $(p,q)=(1,3) \to P_i = 22.693$\,d;
ii)  $(p,q)=(1,2) \to P_i = 25.345$\,d;
iii) $(p,q)=(2,3) \to P_i = 28.699$\,d;
iv)  $(p,q)=(1,1) \to P_i = 39.037$\,d; and
v)   $(p,q)=(3,2) \to P_i = 84.810$\,d. Of these, only options ii) and iii)
appear compatible with the TB prediction with the former leading to a 4:3
resonance between h and i, and the latter a 3:2 resonance (both yield
the same super-period of 489.91\,d). Although technically both are compatible
with the TB law and the Laplacian resonance argument, the 3:2 configuration
- yielding $P_i = 28.699$\,d, would lead to a more generously spaced system.

This work does not address the probability of the existence of an additional
planet (see the work of \citealt{lam:2017} for an example of this). 
However, if an eighth planet is found with one of the two predicted periods,
it would provide some confidence that STIPs with Laplacian resonant chains are
prime systems for precise predictions of planetary periods. 

\acknowledgments

DMK is supported by the Alfred P. Sloan Foundation.


\end{document}